\documentclass[iop]{emulateapj}
\usepackage{apjfonts}
\newcommand{\be}{\begin{equation}}
\newcommand{\ee}{\end{equation}}
\newcommand{\bea}{\begin{eqnarray}}
\newcommand{\eea}{\end{eqnarray}}
\newcommand{\Msun}{\,M_{\odot}}

\shortauthors{CONROY \& BULLOCK}
\shorttitle{Beacons In the Dark}

\begin{document}

\title{Beacons In the Dark: Using Novae and Supernovae to Detect Dwarf
  Galaxies in the Local Universe}

\author{Charlie Conroy\altaffilmark{1} \&
  James S. Bullock\altaffilmark{2}}

\altaffiltext{1}{Department of Astronomy, Harvard University,
  Cambridge, MA, USA}
\altaffiltext{1}{Department of Physics \& Astronomy, University
  of California, Irvine, CA, USA}

\slugcomment{Accepted to ApJ}

\begin{abstract}

  We propose that luminous transients, including novae and supernovae,
  can be used to detect the faintest galaxies in the universe.  Beyond
  a few Mpc, dwarf galaxies with stellar masses $\lesssim10^6\Msun$
  will likely be too faint and/or too low in surface brightness to be
  directly detected in upcoming large area ground-based photometric
  surveys.  However, single epoch LSST photometry will be able to
  detect novae to distances of $\sim30$ Mpc and SNe to Gpc-scale
  distances.  Depending on the form of the stellar mass-halo mass
  relation and the underlying star formation histories of low mass
  dwarfs, the expected nova rates will be a few to $\sim100$ yr$^{-1}$
  and the expected SN rates (including both type Ia and core-collapse)
  will be $\sim10^2-10^4$ within the observable ($4\pi$ sr) volume.
  The transient rate associated with intrahalo stars will be
  comparably large, but these transients will be located close to
  bright galaxies, in contrast to the dwarfs, which should trace the
  underlying large scale structure of the cosmic web.  Aggressive
  follow-up of hostless transients has the potential to uncover the
  predicted enormous population of low mass field dwarf galaxies.

\end{abstract}



\section{Introduction}
\label{s:intro}

Dwarf galaxies are believed to play several important roles in modern
theories of galaxy formation.  The ionizing radiation they emit at
high redshift is likely essential for reionizing the universe
\citep[e.g.,][]{Madau99, Robertson13}.  At later times they contribute
to the buildup of streams and stellar halos around galaxies
\citep[e.g.,][]{Johnston98, Helmi99}.  Their luminosity functions and
kinematic properties place very demanding constraints on models of
star formation and stellar feedback and even perhaps the nature of
dark matter \citep[e.g.,][]{Boylan-Kolchin11, Brooks13, Governato15}.

The relation between stellar mass and dark matter halo mass is a
simple but powerful tool to help understand the distribution of
galaxies in the cosmic web, the luminosity function of dwarf galaxies,
the integrated efficiency of star formation, and the hierarchical
growth of galaxies over cosmic time \citep[e.g.,][]{Purcell07,
  Conroy09d, Behroozi13, Garrison-Kimmel14}.  Empirically-constrained
stellar mass-halo mass relations imply that there exist enormous
numbers of dwarf galaxies with stellar masses $<10^6\Msun$ in the
field \citep[$>1$ Mpc$^{-3}$; e.g., ][]{Behroozi13,
  Garrison-Kimmel14}.  Only a handful of these very low mass field
dwarfs have been detected to date \citep[e.g.,][]{Karachentsev13}.

Their low luminosities and surface brightnesses makes dwarf galaxies
difficult to detect.  Dwarf galaxies within a few Mpc can be resolved
into individual stars.  This has been the classic detection method for
nearby dwarfs in the modern era \citep[e.g.,][]{Willman05, Martin06,
  Belokurov07, Chiboucas09}.  The faintest dwarfs have surface
brightness within a half light radius of $\gtrsim27$ mag arcsec$^{-2}$
\citep{McConnachie12}, making it very difficult to detect them at
distances where they cannot be resolved into stars.  Moreover, there
are theoretical reasons to suspect that a large population of
ultra-diffuse galaxies may exist beyond the detection limits of
current resolved-star surveys \citep[][Wheeler et al., in
preparation]{Bovill09, Bullock10}.  Specialized imagers can reach
these faint surface brightness limits \citep[e.g.,][]{Mihos05,
  Abraham14}, and \citet{Merritt14} recently reported the discovery of
faint dwarf galaxies surrounding M101 detected in this way.  However,
detecting faint dwarfs in the field requires imaging large areas of
the sky, and no such surveys targeting these extremely low surface
brightness limits are currently planned, though clearly they would be
of high value.  At still greater distances dwarfs will appear as faint
point sources, and so either photometric redshifts or follow-up
spectroscopy is required to confirm their dwarf status.

Ongoing and planned high cadence photometric surveys such as the
Palomar Transient Factory \citep[PTF;][]{PTF1,PTF2}, Pan-STARRS
\citep{Kaiser02} the Zwicky Transient Factory \citep[ZTF][]{ZTF1}, and
the Large Synoptic Survey Telescope \citep[LSST;][]{LSST2} are or will
be observing large areas of the sky with the depth necessary to detect
a wide class of transient objects in single epoch imaging and
individual bright stars in nearby dwarf galaxies in coadded images.
In this Letter we propose that the transients that will be discovered
in wide field photometric surveys can be used to detect faint dwarf
galaxies that would otherwise be too distant to detect with standard
techniques.  This proposal is similar in spirit to the idea of using
transients to detect the diffuse population of stars in the halos of
galaxies, groups, and clusters, often known as intrahalo stars
\citep[e.g.,][]{Gal-Yam03, Shara06, Sand12}.  \citet{Shara06} for
example predicted that LSST will observe hundreds of novae associated
with intrahalo stars.


\section{Dwarf Galaxies In the Local Group}

\begin{figure}[!t]
\center
\includegraphics[width=0.47\textwidth]{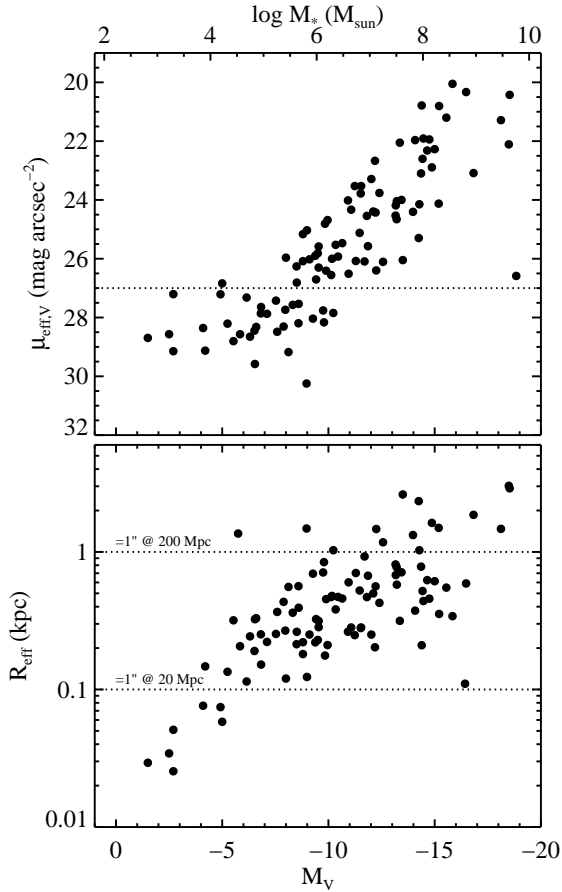}
\caption{Properties of known dwarfs in the Local Group, from data
  compiled by \citet{McConnachie12}.  The dotted line in the upper
  panel marks the rough expected surface brightness limit of LSST.
  Stellar masses have been estimated assuming $M/L_{\rm V}=2.0$.}
\label{fig:dwarfs}
\end{figure}

\citet{McConnachie12} provided a compilation of key photometric and
structural paramaters of all known dwarf galaxies (as of 2012) within
3 Mpc of the Sun.  From the reported effective radii, magnitudes, and
ellipticities we have computed the mean $V-$band surface brightness
within the effective radius, $\mu_{\rm eff,V}$.  The distribution of
known Local Group dwarf galaxies in size-luminosity-surface brightness
space is shown in Figure \ref{fig:dwarfs}.  We also translate the
sizes into distances at which the size would subtend $1\arcsec$.  A
threshold surface brightness of $\mu_{\rm eff,V}=27$ mag arcsec$^{-2}$
is indicated as a rough estimate of the LSST surface brightness limit
\citep{LSST2}.  Notice that the faintest known galaxies all lie at
$\mu_{\rm eff,V}>27$ mag arcsec$^{-2}$, suggesting that it will be
difficult for LSST to detect such galaxies except where the individual
stars from such galaxies can be easily detected (e.g., within a few
Mpc of the Sun).  In the figure we also translate the x-axis into an
approximate stellar mass by assuming a $V-$band mass-to-light ratio of
$M/L_{\rm V}=2.0$, appropriate for old (10 Gyr) metal-poor
([Z/H]$<-1$) populations \citep{Conroy09a}\footnote{A \citet{Kroupa01}
  stellar initial mass function has been assumed throughout this work
  when quoting stellar masses and rates per unit stellar mass.}.  Of
course, for the star-forming galaxies this is a poor assumption and
will result in an overestimation of the mass.  These masses are only
computed to guide the eye and are not used in any of the calculations
below.

\vspace{2cm}

\section{Model Ingredients}
\label{s:model}

In this section we present a model for the expected SN and nova rates
from dwarf galaxies and intrahalo stars.

We start with the $z=0$ dark matter halo mass function from
\citet{Tinker08}.  The halo virial mass is defined according to
\citet{Bryan98}.  We adopt a Hubble constant of $H_0=70$ km s$^{-1}$
Mpc$^{-1}$, $\Omega_m=0.27=1-\Omega_\Lambda$, and $\sigma_8=0.8$,
consistent with the 7-yr WMAP results \citep{WMAP7}.  These parameters
are somewhat different from the recent Planck measurements
\citep{Planck15}, but the differences have no material impact on our
results.

\begin{figure}[!t]
\center
\includegraphics[width=0.54\textwidth]{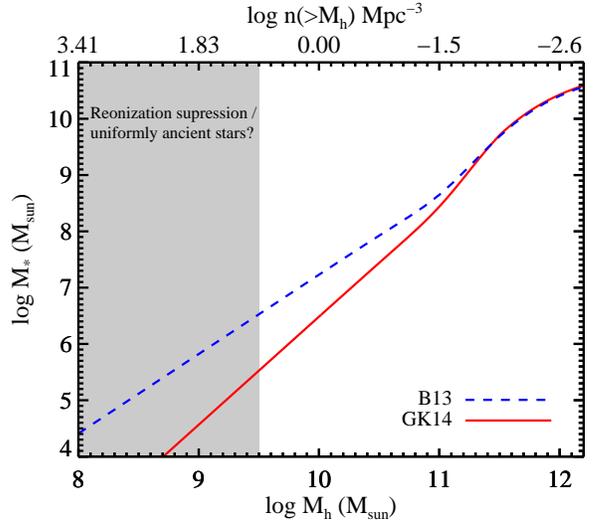}
\caption{Stellar mass-halo mass relations.  The upper $x-$axis has
  been converted to cumulative number density based on our fiducial
  cosmology.  The B13 relation is a significant extrapolation beyond
  the mass range over which the original relation was constrained.
  The shaded region marks the approximate region where reionization is
  expected to quench star formation in low mass halos.}
\label{fig:smhm}
\end{figure}

Galaxies are placed within dark matter halos according to a stellar
mass-halo mass ($M_\ast-M_h$) relation.  Our fiducial relation is
adopted from \citet[][GK14]{Garrison-Kimmel14}, which reproduces both
the empirical galaxy stellar mass function at $z=0$ and the Local
Group dwarf luminosity function.  The form of the $M_\ast-M_h$
relation at low masses is uncertain and has a large effect on our
model predictions.  In order to explore this sensitivity we will also
consider the relation from \citet[][B13]{Behroozi13}, which we
extrapolate well beyond the mass range probed by B13 (roughly
$M_\ast=10^8-10^{11}\Msun$).  The two relations are compared in Figure
\ref{fig:smhm}.  We also indicate with a shaded band the approximate
region where reionization is expected to quench star formation in low
mass halos \citep[e.g.,][]{Bullock00}.  We adopt a relation with no
scatter whereas in reality there may in fact be significant scatter at
low masses \citep[][Garrison-Kimmel et al., in preparation]{Wang15}.
As long as the scatter is symmetric about the mean relation this will
not affect our predicted rates.

\begin{figure*}[!t]
\center
\includegraphics[width=0.49\textwidth]{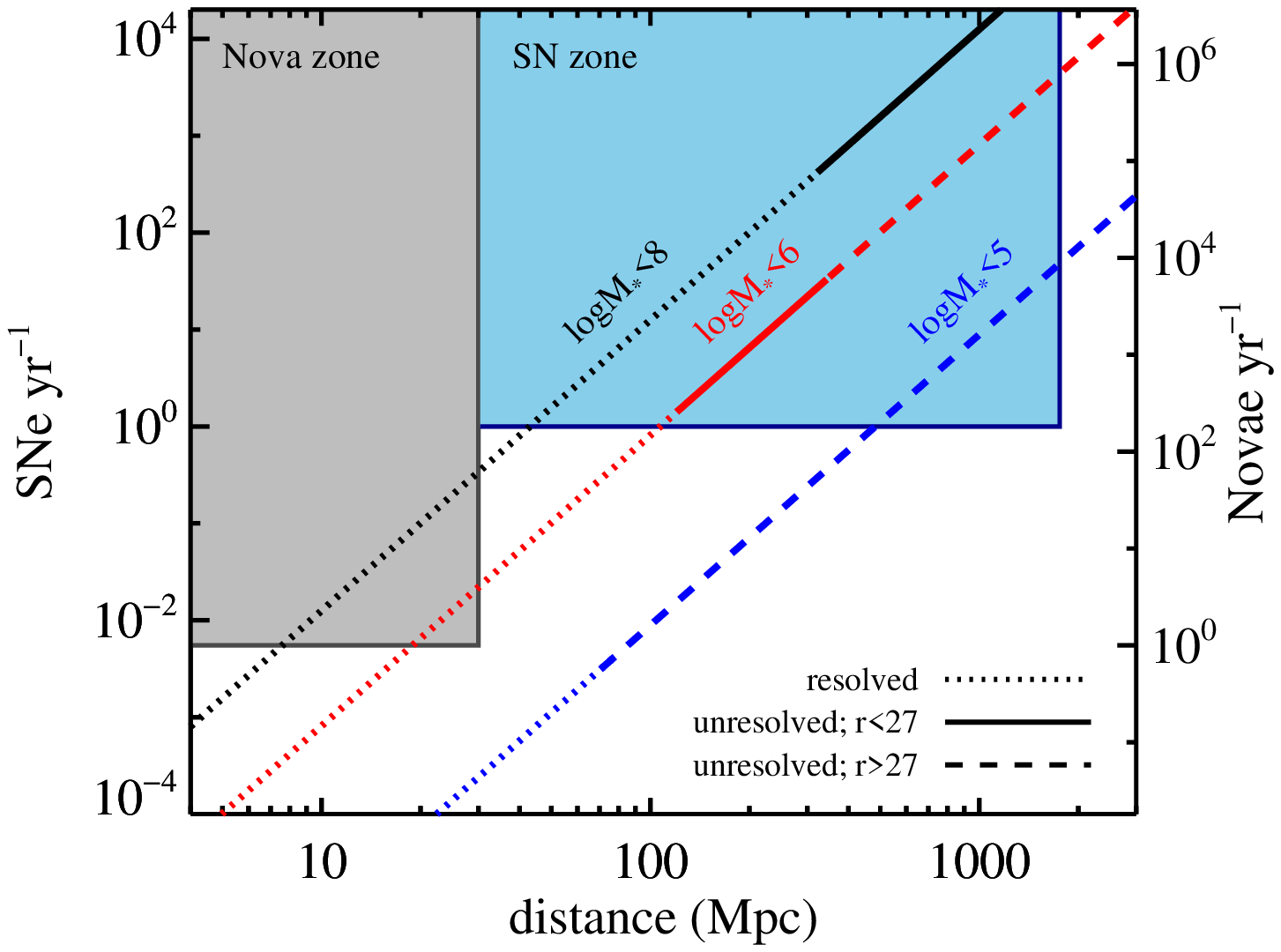}
\includegraphics[width=0.49\textwidth]{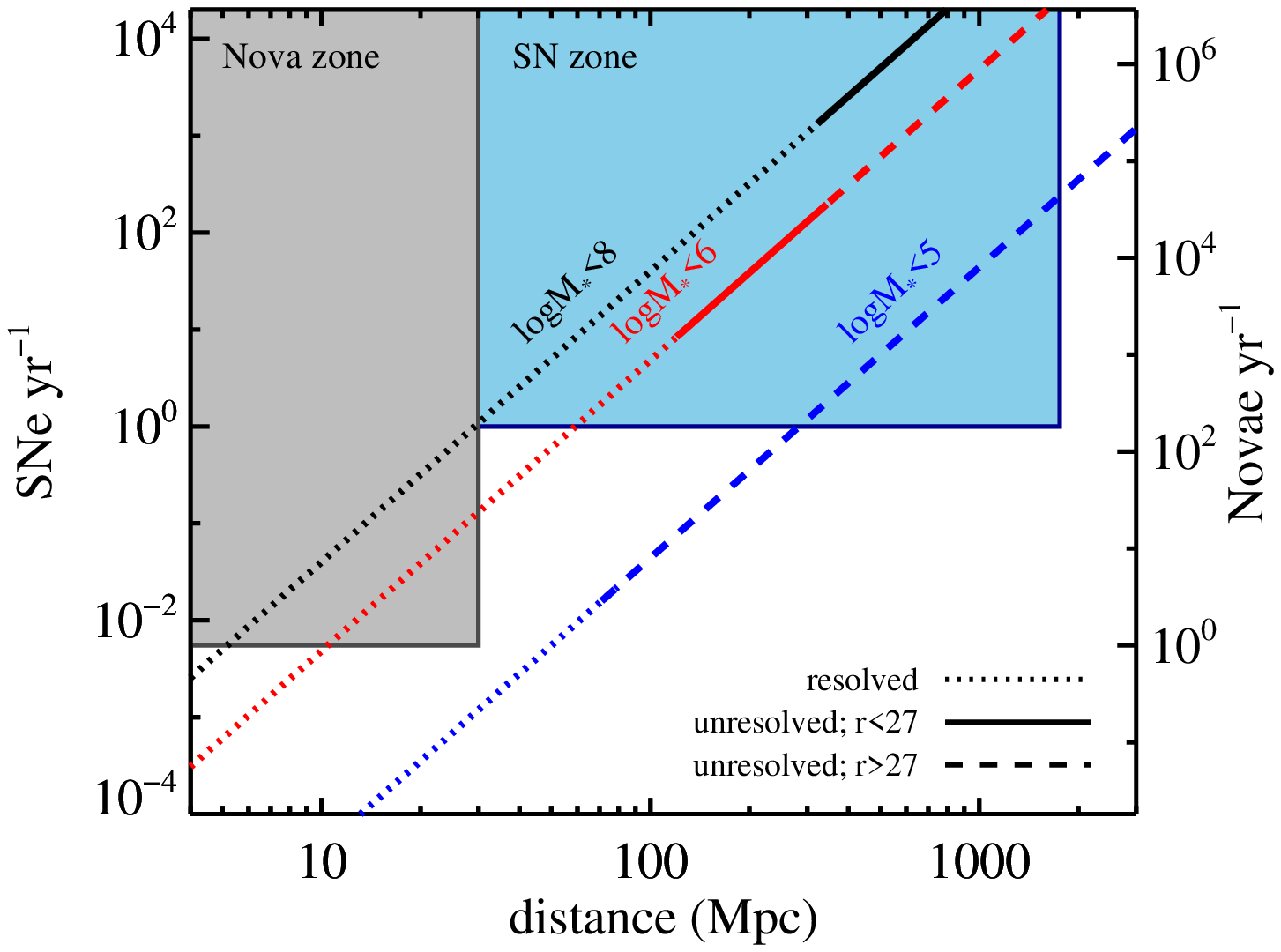}
\caption{Predicted SN and nova rates as a function of distance
  adopting the GK14 (left) and B13 (right) stellar mass-halo mass
  relations.  Lines show rates for dwarf galaxies with stellar masses
  less than $10^8\Msun$, $10^6\Msun$, and $10^5\Msun$. The shaded
  regions indicate where SNe and novae are detectable with a single
  LSST visit ($r<24.3$) and where the rates are $>1$ yr$^{-1}$.  Each
  line is broken into three regions: where a typical galaxy would be
  spatially resolved (dotted), unresolved and brighter than $r=27$
  (solid) and both unresolved and fainter than $r=27$ (dashed).  This
  depth was chosen to be comparable to the final 10 yr LSST coadded
  depth of $r\approx27.5$.  Note that the SN/nova ratio is different
  for the lowest mass bin compared to the higher mass bins owing to
  the different assumed SFHs.  The quoted nova rates on the $y-$axes
  are therefore only appropriate for the two most massive bins.}
\label{fig:rates}
\end{figure*}

The next step is to specify star formation histories (SFHs) for the
dwarfs, as the SFHs will influence the SN rates.  We adopt the
following simple model: dwarf galaxies with $M_\ast>10^5\Msun$ are
assumed to have a constant SFR throughout their lifetimes, while less
massive dwarfs are assumed to form all of their stars instantly at
high redshift ($z>2$).  We consider in this work only dwarfs in the
field, i.e., dwarfs that are not within the virial radius of a larger
halo.  This model is motivated by SFHs derived from resolved color
magnitude diagrams of Local Group dwarfs \citep[e.g.,][]{Weisz14} and
also consistent with expectations for dwarfs above and below this
stellar mass threshold in cosmological simulations
\citep[e.g.,][Wheeler et al. in preparation]{Onorbe15}.  Later, when
converting masses into $r-$band magnitudes, we adopt $M/L_r=1.8$ for
the low mass galaxies with ancient metal-poor stellar populations and
$M/L_r=0.73$ for the higher mass dwarfs with constant SFHs
\citep{Conroy09a}.

We will generically refer to SNe, noting that this includes both the
core-collapse (CC) and type Ia varieties.  For the CC SN rate we
assume 1 per $100\Msun$ of newly formed stars, i.e., a galaxy with a
star formation rate of $1\Msun$ yr$^{-1}$ will have CC SN rate of
$10^{-2}$ yr$^{-1}$.  It is well-known that the rate of Ia SN depends
on the age of the progenitor stellar population, with a delay time
distribution (DTD) that falls approximately as $t^{-1}$.  We adopt a
DTD based on data presented in \citet{Maoz12}, with a rate at 10 Gyr
of $3\times10^{-14}$ yr$^{-1}$ $\Msun^{-1}$ and a time dependence of
$t^{-1}$.  With this DTD, a galaxy with a constant SFH will at the
present epoch have a Ia rate of $\approx10^{-13}$ yr$^{-1}$
$\Msun^{-1}$ (assuming the DTD is zero at $t<10^8$ yr).  The above
assumptions imply that the CC rate is $10\times$ higher than the Ia
rate for a galaxy with a constant SFH.

Distances to the SNe are required to use them to detect low redshift
dwarf galaxies.  For type Ia SNe, deriving distances from the light
curve is straightforward.  There has been some effort to use a subset
of CC SNe (the ``plateau'' type) as standard candles
\citep[e.g.,][]{Schmidt94, Hamuy02}.  \citet{Hamuy02} show a
reasonably strong correlation between type IIP SNe apparent magnitude
and redshift, which should be sufficient to at least discriminate
between high and low redshift SNe.  Spectroscopic follow-up of
transients, which will no doubt occur for a subset of the events, will
also be valuable for confirming the redshift of these apparently
hostless SNe and therefore also the redshift of the underlying faint
dwarf galaxy.

Nova rates have historically been difficult to empirically estimate
owing to a variety of selection and incompleteness effects.
Nonetheless, observations of galaxies in the nearby universe are
broadly consistent with a universal nova rate of $2\times10^{-10}$
yr$^{-1}\,\Msun^{-1}$, independent of Hubble type, galaxy color,
etc. \citep{Ciardullo90, Ferrarese03, Williams04}.  There is however
some evidence that the nova rate may increase toward lower mass
galaxies \citep{Neill05}, which, if confirmed, would have important
implications for the present study.  Novae peak absolute magnitudes
fall in the range from $-7\lesssim M_V\lesssim-9$; in this work we
adopt a typical peak magnitude of $M_r=-8$.  The decay time of the
nova light curve is inversely related to its peak brightness and the
relation is sufficiently tight, and theoretically well-understood, so
that one may be able to infer distances to novae from their light
curves \citep[e.g.,][]{Shara81, Downes00}.

\citet{Purcell07} presented a simple model for the buildup of
intrahalo light (IHL), where intrahalo stars are defined as residing
within a parent dark matter halo but not bound to any particular
galaxy.  The model assigns stars to dark matter halos in a
cosmological $N-$body simulation via an empirically-constrained
$M_\ast-M_h$ relation, and when a dark matter subhalo is disrupted in
the simulation the stars from that subhalo are assigned to the IHL.
At the cluster scale, this model produces IHL fractions of order unity
while at the Milky Way halo scale the IHL contributions are at the
percent level; both of these predictions are in agreement with
observations \citep[e.g.,][]{Gonzalez05, Irwin05}.  We have fit their
results with a polynomial function:
\noindent
\be
f_{\rm IHL}(M_h)\equiv \frac{M_{\rm \ast,IHL}}{M_{\rm \ast,gal}}=\sum_{i=0}^3\, a_i\, ({\rm log}\,M_h)^i, 
\ee 
\noindent
with $a_i=(9.5936,-5.3943,0.5965,-0.01869)$, where $M_{\rm \ast,IHL}$
and $M_{\rm \ast,gal}$ are the stellar masses in the IHL and the
central galaxy, respectively.  The fit reproduces the model results to
$5-10$\%.  We assume that the intrahalo stars are uniformly old when
computing nova and SN rates from this component.  Integrating over
the halo mass function, the above equation implies that the total
stellar mass in the IHL is $\approx10$\% of the stellar mass in galaxies.

\vspace{0.2cm}

\section{Results}
\label{s:results}

The resulting model SN and nova rates for dwarf galaxies are shown in
Figure \ref{fig:rates} as a function of distance.  We show results for
two stellar mass-halo mass relations in order to highlight the
sensitivity to the adopted relation.  Clearly, observations of the SN
and nova rate could place powerful constraints on the low mass end of
the $M_\ast-M_h$ relation.  Rates are shown for dwarfs with stellar
masses less than $10^8\Msun$, $10^6\Msun$, and $10^5\Msun$.  Along
each line we mark where the most massive dwarfs would be
resolved\footnote{Here we use ``resolved'' and ``unresolved'' to refer
  to spatially resolving the galaxy, as opposed to resolving the
  galaxy into individual stars.  We consider a galaxy ``resolved'' if
  the effective radius is $>0\arcsec.5$.}, unresolved and brighter
than $r=27$, and unresolved and fainter than $r=27$.  This depth was
chosen to be comparable to the final, 10 yr coadded LSST depth of
$r\approx27.5$ \citep{LSST2}.  In order to compute when a dwarf would
be resolved we have assumed effective radii of 0.8, 0.3, and 0.2 kpc
for dwarfs with stellar masses of $10^8\Msun$, $10^6\Msun$, and
$10^5\Msun$, respectively (see Fig \ref{fig:dwarfs}).  Shaded regions
denote ``sweet spots'' where the rates are $>1$ yr$^{-1}$ and the
transients are observable.  In the case of novae the maximum distance
is set by being able to detect a $M_r=-8$ nova in single epoch LSST
data (though of course multiple epochs will be required to confirm
that it is a nova), which will have a minimum depth of $r\approx24.3$
\citep{LSST2}.  In the case of SNe the maximum depth is set by being
able to detect a $M_r=-18$ SN in single epoch LSST data.  This is a
conservative limit as both type Ia and CC SNe can have peak magnitudes
brighter than this.  Finally, note that the quoted rates are for the
entire ($4\pi$ sr) observable volume, whereas in practice any survey
will cover only a fraction of the sky at the necessary cadence.

The principle result of this Letter is that SNe and novae residing in
low mass dwarf galaxies should be detectable in large numbers with
upcoming time domain surveys such as LSST.  Dwarfs in the mass range
of $M_\ast\sim10^5-10^6\Msun$ should be spatially resolvable to
$10-100$ Mpc distances, but their low surface brightness, $\mu>26$ mag
arcsec$^{-2}$, will make it challenging to directly detect them.
However, novae within these low mass galaxies will be detected at
rates of $1-10^2$ yr$^{-1}$ within $\sim30$ Mpc, and SNe will be
detected out to Gpc scales at rates of $\sim10^2-10^4$ yr$^{-1}$ (note
that a comoving radial distance of 1 Gpc corresponds to $z=0.25$ in
our adopted cosmology).  Even very low mass dwarfs with
$M_\ast<10^5\Msun$ will produce type Ia SNe at a rate of $\sim10^2$
yr$^{-1}$.  Relatively massive dwarfs ($M_\ast\sim10^8\Msun$), should
be easily detectable as spatially resolved, relatively high surface
brightness objects out to several hundred Mpc distances.  For these
objects, detection via transients will likely be less interesting.

\begin{figure}[!t]
\center
\includegraphics[width=0.49\textwidth]{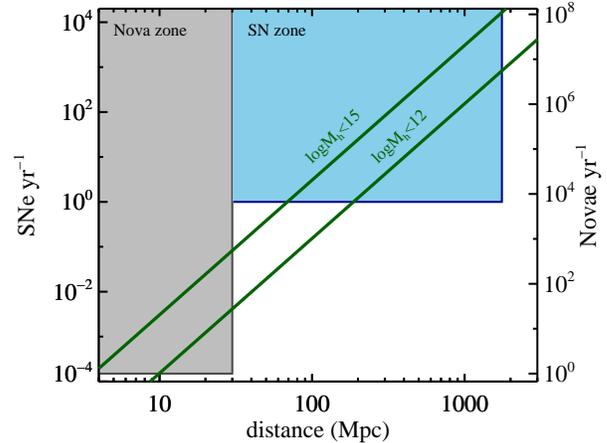}
\caption{Same as Figure \ref{fig:rates}, except now for intrahalo
  stars residing in halos less massive than $10^{15}\Msun$ and
  $10^{12}\Msun$.  The intrahalo stars are assumed to be uniformly
  ancient and so the SN/nova ratio is different here than in the
  $y-$axes in Figure \ref{fig:rates}.}
\label{fig:ihl}
\end{figure}

Figure \ref{fig:ihl} shows the expected SN and nova rates from
intrahalo stars.  Results are shown for stars within dark matter halos
less massive than $10^{15}\Msun$ and $10^{12}\Msun$.  The implied nova
rates within 30 Mpc are fairly high, reaching nearly $\approx500$
yr$^{-1}$ if one considers all IHL stars.  A key difference between
the transients associated with IHL stars and dwarfs is their spatial
distribution - IHL stars are confined to within $0.1-1$ Mpc (depending
on the virial radius of the host halo), while the dwarfs are
presumably distributed throughout space, tracing the underlying dark
matter distribution with a constant anti-bias.  The implied IHL rates
are broadly consistent with, although somewhat lower than the
predictions from \citet{Shara06}.  Shara assumed a global IHL
fraction, $f_{\rm IHL}$, of $0.1-0.4$, whereas our model for the IHL,
constrained by recent data, places this fraction at closer to 0.1.

\vspace{0.2cm}

\section{Discussion}
\label{s:disc}

Dwarfs at the mass scale $M_\ast\sim10^5-10^6\Msun$ reside within some
of the smallest dark matter halos that are expected to form stars -
scales where the physics of reionization and dark matter may be probed
with potentially powerful effect.  Unfortunately, their use as cosmic
laboratories to this point has been fundamentally limited by the fact
that we detect them with reasonable completeness only within $\sim$Mpc
volumes.  As Figure \ref{fig:rates} makes clear, the detection of
(apparently) hostless SNe at $0.1-2$ Gpc distances should provide an
important avenue for detecting these dwarfs and constraining the
$M_\ast-M_h$ relation at low masses within volumes that are virtually
immune to sample variance effects.  In addition to counts, the ratio
of type Ia to CC SNe will be a sensitive probe of the average SFH of
the unresolved dwarfs.  The correlation function of hostless SNe
within a Gpc will provide constraints on the halo masses of the dwarfs
and will allow for a statistical separation of transients arising from
dwarfs and from intrahalo stars.  One could also imagine analyzing the
rates as a function of large scale environment to probe the low mass
end of the galaxy population in e.g., void regions.

The majority of the dwarf galaxies detected via transients will be too
distant for detailed study.  We envision this detection technique as
primarily a statistical probe of the low mass dwarf population.  A
dwarf with $M_\ast=10^6\Msun$ and a constant SFH will have a CC SN
rate of $10^{-6}$ yr$^{-1}$, so one must observe a volume containing
$10^6$ such dwarfs to detect one SN per year.  This technique is only
feasible because of the enormous numbers of predicted dwarf galaxies
in the field.

A number of assumptions were made in computing the rate of luminous
transients, the most important of which is the adopted $M_\ast-M_h$
relation.  Other important assumptions include the SFH of the dwarfs
and the assumed constant nova rate per unit stellar mass.  We have
assumed that galaxies with $M_\ast>10^5\Msun$ have constant SFHs in
the field.  If instead some fraction are quenched or the star
formation rates were higher in the past then the predicted SN rates
will be lower.  On the other hand, if the nova rate is in fact higher
in dwarf galaxies, as suggested by \citet{Neill05}, then the expected
nova rate from dwarf galaxies could be much higher than assumed here.
LSST and other transient surveys will directly constrain the nova rate
in nearby dwarf galaxies with well-studied stellar populations, which
will effectively eliminate the nova rate as a source of uncertainty.
We have also assumed that the CC SN rate per unit star formation rate
is constant and does not change with metallicity nor with the level of
star formation rate.  The latter in particular might affect the SN
rate at levels where stochasticity and/or the effects of small cloud
masses become important \citep[e.g.,][]{Weidner04, daSilva12}.  Here
again LSST and other transient surveys will provide rates of CC SNe
within known galaxies to inform a better model for the expected CC SN
rate in low mass dwarfs.  Finally, we have included CC SNe in the
transient budget, but it is not clear if these will be as useful as
novae and type Ia SNe due to the difficulty in assigning distances to
CC SNe based only on the light curve.  However, even in the most
pessimistic case in which one discards the CC SN events, the predicted
Ia rates for $M_\ast>10^5\Msun$ are still substantial (approximately
$10\times$ lower than shown in Figure \ref{fig:rates}).  Moreover,
recall that our model for the lowest mass galaxies
($M_\ast<10^5\Msun$) assumes an ancient stellar population and hence
the SNe associated with those galaxies are solely of the Ia variety.


\acknowledgments 

We thank Jeremy Tinker for providing code to compute halo mass
functions and Mike Boylan-Kolchin, Mansi Kasliwal, Pieter van Dokkum,
and Dan Weisz for comments on an earlier draft.  CC is supported by a
Packard Foundation Fellowship.  This research was supported in part by
the National Science Foundation under Grant No. NSF PHY11-25915.  The
authors acknowledge the KITP workshop "Galactic Archaeology and
Precision Stellar Astrophysics" where this work was conceived.  Ned
Wright's online cosmology calculator \citep{Wright06} was used in the
preparation of this manuscript.

\end{document}